\documentclass[reprint,superscriptaddress,amsmath,amssymb,aps]{revtex4-1}
\usepackage{graphicx}% Include figure files
\usepackage{dcolumn}% Align table columns on decimal point
\usepackage{bm}% bold math 
\usepackage{graphicx} 
\usepackage{mathrsfs}
\usepackage{mathtools}
\usepackage{balance}
\usepackage{hyperref}
\usepackage[USenglish]{babel}

\begin{document}

\title{Quantum--continuum simulation of the electrochemical response \\ of pseudocapacitor electrodes under realistic conditions}

\author{Nathan Keilbart}
\email{Email:nathan.keilbart@psu.edu}
\affiliation{Department of Materials Science and Engineering, Materials Research Institute, and Penn State Institutes of Energy and the Environment, The Pennsylvania State University, University Park, PA 16802, USA}
\author{Yasuaki Okada}
\affiliation{Department of Materials Science and Engineering, Materials Research Institute, and Penn State Institutes of Energy and the Environment, The Pennsylvania State University, University Park, PA 16802, USA}
\affiliation{Murata Manufacturing Co., Ltd., 10-1, Higashikotari 1-chome, \\ Nagaokakyo-shi, Kyoto 617-8555, Japan}
\author{Aion Feehan}
\affiliation{Department of Materials Science and Engineering, Materials Research Institute, and Penn State Institutes of Energy and the Environment, The Pennsylvania State University, University Park, PA 16802, USA}
\affiliation{{\' E}cole Centrale Paris, Grande Voie des Vignes, \\ 92290 Ch{\^ a}tenay-Malabry Cedex, France}
\author{Shinichi Higai}
\affiliation{Murata Manufacturing Co., Ltd., 10-1, Higashikotari 1-chome, \\ Nagaokakyo-shi, Kyoto 617-8555, Japan}
\author{Ismaila Dabo}
\affiliation{Department of Materials Science and Engineering, Materials Research Institute, and Penn State Institutes of Energy and the Environment, The Pennsylvania State University, University Park, PA 16802, USA}

\begin{abstract}
Pseudocapacitors are energy-storage devices characterized by fast and reversible redox reactions that enable them to store large amounts of electrical energy at high rates. We simulate the response of pseudocapacitive electrodes under realistic conditions to identify the microscopic factors that determine their performance, focusing on ruthenia (RuO$_2$) as a prototypical electrode material. Electronic-structure methods are used together with a self-consistent continuum solvation (SCCS) model to build a complete dataset of free energies as the surface of the charged electrode is gradually covered with protons under applied voltage. The resulting dataset is exploited to compute hydrogen-adsorption isotherms and charge--voltage responses by means of grand-canonical sampling, finding close agreement with experimental voltammetry. These simulations reveal that small changes on the order of 5 $\mu$F/cm$^2$ in the intrinsic double-layer capacitance of the electrode--electrolyte interface can induce variations of up to 40 $\mu$F/cm$^2$ in the overall pseudocapacitance.
\end{abstract}

\maketitle

\section{\label{sec:Introduction}Introduction}

\begin{figure}[t!]
	\centering
	\includegraphics[width=\columnwidth]{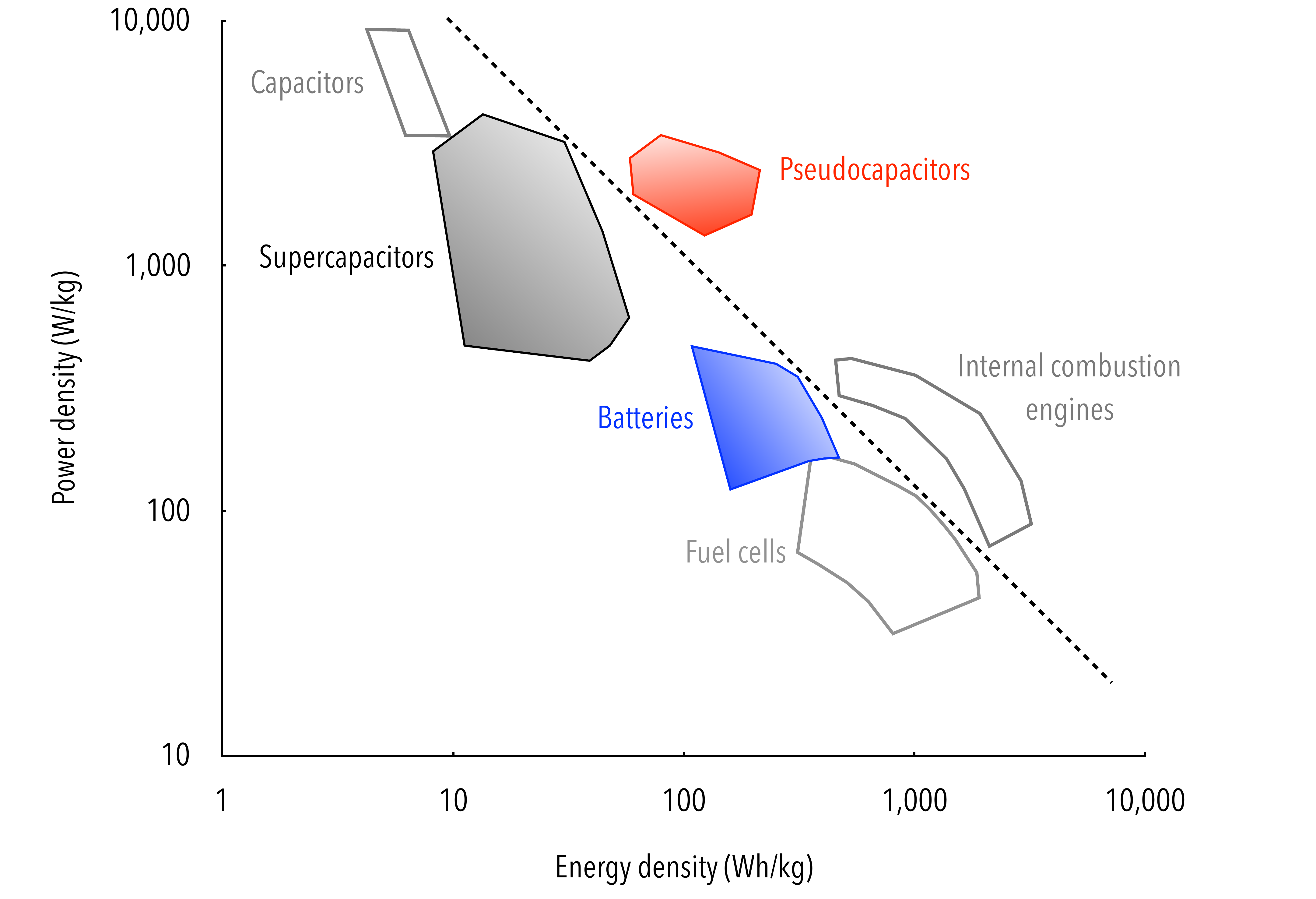}
	\caption{Pseudocapacitors are energy-storage devices combining the advantages of batteries and supercapacitors; they aim to bring the energy-storage performance of electrochemical systems closer to that of internal combustion engines.
		\label{fig:electrochemical-frontier}}
\end{figure}

Electrochemical energy storage is dominated by batteries and supercapacitors; batteries exhibit high energy capacities but low charging rates, whereas supercapacitors are characterized by fast charging times but low energy densities. The development of advanced technologies combining the energy density of batteries with the power density of supercapacitors is critical to overcome the frontier separating the performance of electrochemical systems from that of internal combustion engines and meet the technical requirements for electric transportation and grid energy storage, as illustrated in Fig.~\ref{fig:electrochemical-frontier}.

Ruthenia (RuO$_2$) is known to be highly efficient at storing large amounts of energy on time scales comparable to those of supercapacitors \cite{Trasatti1971, Simon2008, Gonzalez2016, Augustyn2014, Bandaru2015, Ardizzone1990} due to fast and reversible interfacial redox processes known as pseudocapacitive reactions \cite{Conway1991, Mckeown1999, Sugimoto2006, Wang2016,  Liu2012, Ozolins2013, Over2012}. Nevertheless, the high cost of ruthenium is a major hurdle to the commercial deployment of ruthenia-based  devices.  These strong constraints motivate the search for low-cost materials whose performance will be comparable to that of RuO$_2$ electrodes.

\begin{figure}[t!]
	\centering
	\includegraphics[width=\columnwidth]{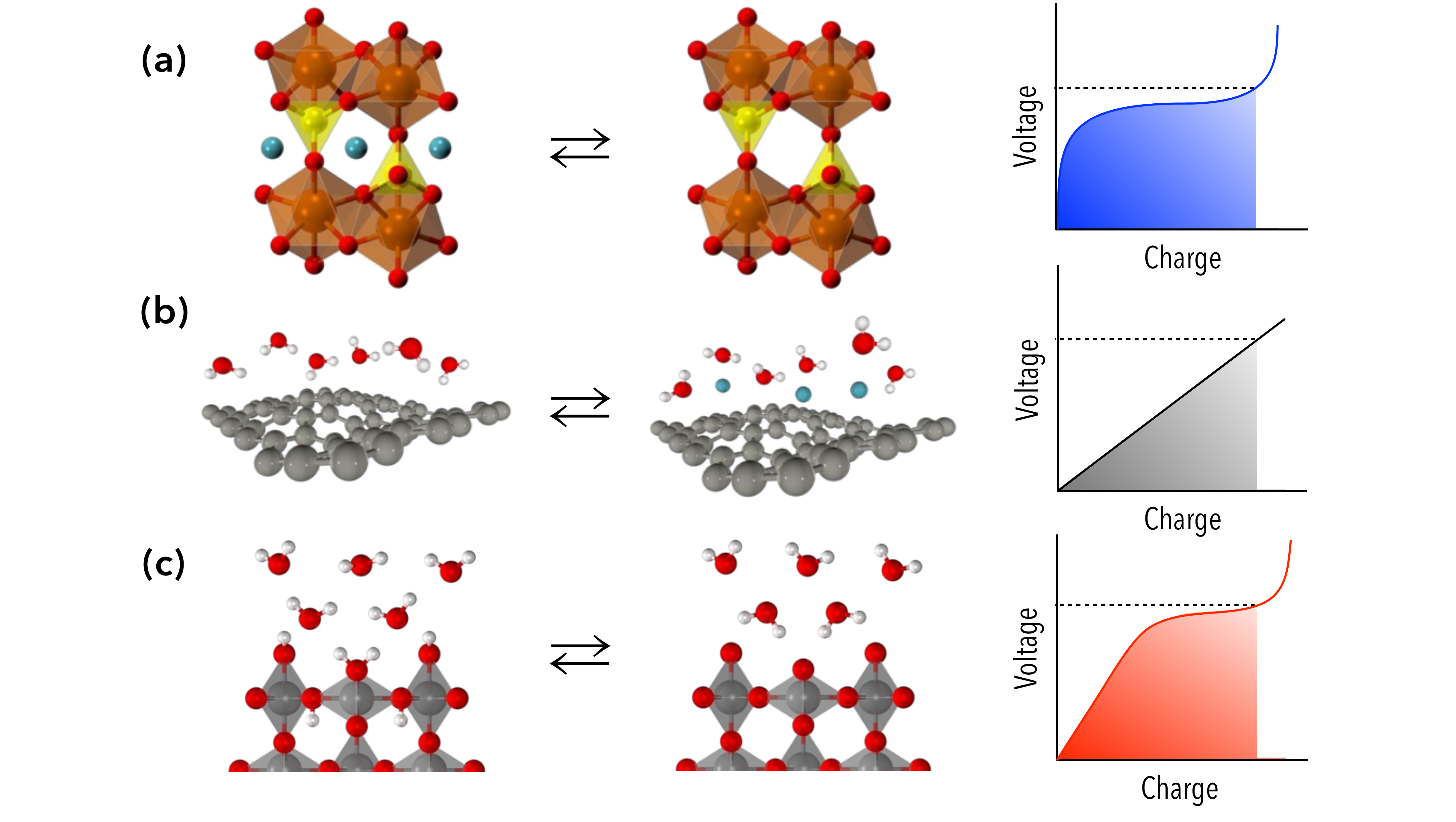}
	\caption{The energy-storage capacity of (a) batteries, (b) supercapacitors, and (c) pseudocapacitors corresponds to the area below their charge--voltage response up to the maximal voltage determining the stability of the electrochemical cell (dashed line).
		\label{fig:charge-voltage-response}}
\end{figure}

Recently, there has been considerable progress in predicting energy storage in electrode materials \cite{Ceder2012, Urban2016}.  As illustrated in Fig.~\ref{fig:charge-voltage-response}(a), these calculations consist of simulating the insertion of ions into the crystal structure of the electrode to determine the electrochemical energy stored during the intercalation process \cite{Zhou2004,Wolverton1998}. Similar methods can be used to simulate capacitive and pseudocapacitive energy storage provided that the complexity of the oxide--solution interfacial environment is taken into account. In specific terms, studying pseudocapacitive processes requires one to consider the time-dependent evolution of the electrolyte at the electrified interface, as depicted in Figs.~\ref{fig:charge-voltage-response}(b) and \ref{fig:charge-voltage-response}(c).  Although this problem can be tackled by leveraging the accuracy of first-principles molecular dynamics, these simulations are computationally demanding, making it challenging to apply them to the combinatorial discovery for pseudocapacitive oxides. 

Consequently, it is necessary to develop scalable computational approaches that will capture the detailed features of electrified solid--liquid interfaces at tractable computational cost \cite{Wood2014, Karlberg2007, Bonnet2013}.  We address this problem by building and validating a comprehensive model of the solid--solution interface under realistic electrochemical conditions.  This model extends the capabilities of first-principles methods in predicting the energy-storage capacity of pseudocapacitive electrodes by integrating a quantum-mechanical description of the electrode covered by adsorbed species with a continuum representation of the surrounding electrolyte.  We employ this model to build an exhaustive dataset of solvated equilibrium structures for protons adsorbed at the surface of the electrode and carry out large-scale Monte Carlo simulations under applied voltage for predicting the charge--voltage response and electrical performance of the interface at finite temperature.  In the following, we outline the method used in the quantum--continuum modeling of ruthenia--electrolyte interfaces and the large-scale simulation of their electrochemical response. We then present our computational predictions and discuss their implications in optimizing energy storage in pseudocapacitors.

%%%%%%%%%%%%%%%%%%%%%%%%%%
\section{Computational methods}
%%%%%%%%%%%%%%%%%%%%%%%%%%

\label{sec:computational_methods}

\subsection{Quantum--continuum modeling}

Density-functional theory (DFT) calculations are performed with the {\sc pw} code of the Quantum-Espresso distribution \cite{Giannozzi2009}. The Perdew--Burke--Ernzerhof (PBE) \cite{Perdew1996} exchange-correlation functional is employed with ultrasoft pseudopotentials to represent atomic cores. Kinetic energy cutoffs of 50 Ry and 500 Ry are used for the plane-wave  expansion of the wave functions and charge density, respectively. The bulk crystal structure of RuO$_2$ is calculated through variable cell relaxation. The Brillouin zone is sampled with a grid of $4 \times 4 \times 6$ points and the electronic occupations are calculated with 0.03 Ry of Marzari--Vanderbilt cold smearing. It is important to note that the description of strong electronic correlations in transition metal oxides typically involves the use of advanced electronic-structure methods such as the onsite Hubbard correction, beyond conventional local and semilocal density functionals \cite{Cococcioni2005}; however, in the case of rutile RuO2, the Hubbard correction is not necessary due to the well-known metallic nature of this electrode material  \cite{Liu2012,Ozolins2013,Rossmeisl2007,Sun2003,Watanabe2016,Wang2007}. After variable cell relaxation, we obtain values of $a$ = 4.64 \AA\ and $c$ = 3.19 \AA, reflecting the slight tendency of the semilocal functional to overestimate the experimental lattice parameters, $a$ = 4.50 \AA\ and $c$ = 3.10 \AA\ \cite{Bolzan1997}.  We employ the calculated parameters to construct our solvated slab models. 

In these calculations, environmental solvent effects are included by means of the self-consistent continuum solvation (SCCS) method --- a continuum model that has been developed to be transferable between molecular systems and solvated surfaces, and that has been parameterized to reproduce the electrical potential of the electrodes in their neutral and charged states \cite{Andreussi2012}. In this approach, a dielectric cavity is constructed at the surface of the system. The dielectric permittivity is taken to be a self-consistent function of the electron charge density; it is written as $\epsilon = \exp\left[  {\left(\zeta-{\sin(2\pi \zeta)}/{2\pi}\right)}  \ln \epsilon_0 \right]$, where $\epsilon_0$ is the dielectric  constant of the bulk solvent and the variable $\zeta$ is defined as $\zeta=({\ln \rho_{\rm max}-\ln\rho})/({\ln\rho_{\rm max}-\ln\rho_{\rm min}})$ with $\rho_{\rm max}$ and $\rho_{\rm min}$ being the density thresholds that delimit the internal and external isocontours of the smooth dielectric cavity. This model also includes non-electrostatic effects such as the external pressure and surface tension, as well as dispersion and repulsion effects. Explicitly, these contributions are written as $G_{\rm cav}=\gamma S$ and $G_{\rm dis+rep}=\alpha S+\beta V$, where $\gamma$ is the experimental solvent surface tension, and $\alpha$ and $\beta$ are fitted parameters. $S$ and $V$ are the quantum surface and volume for the solute that are defined as $S=-\int d\textbf{r}\frac{d\theta}{d\rho}(\rho(\textbf{r}))|\nabla\rho(\textbf{r}) |$ and $V=\int d\textbf{r}\theta(\rho(\textbf{r}))$, which involve the smooth switching function $\theta(\rho)=(\epsilon_0-\epsilon(\rho))/(\epsilon_0-1)$.  In specific terms, we employ the following parameterization of the solvent, which has been extensively fitted on more than 240 molecules and found to be in good agreement with both experiment and the widely used polarizable continuum model (PCM) as implemented in Gaussian09 \cite{Andreussi2012}.  This model has then been further refined to provide accurate solvation energies for either charged molecular anions or molecular cations \cite{Dupont2013} and accurate electrode potentials for solvated metal electrodes \cite{Weitzner2017}: $\epsilon_0 = 78.3$, $\rho_{\rm min}=10^{-4}$ a.u., $\rho_{\rm max}=5 \times 10^{-3}$ a.u., $\gamma = 72.0$ dyn/cm, $\alpha = -22$ dyn/cm, and $\beta = -0.35$ GPa. 

\subsection{Supercell surface structures}

\label{sec:surface-structures}

We apply the quantum--continuum model to simulate RuO$_2$ surfaces. A review of the literature indicates that the RuO$_2$(110), (100), and (101) orientations are all pseudocapacitive, but that the (110) crystallographic plane is one of the most active and best characterized \cite{Lister2002,Ozolins2013,Liu2012,Guerrini2005}.  We thus focus our attention on this surface orientation. A supercell slab model of the (110) facet is created with a $2 \times 1$ unit cell allowing us to correctly account for first- and second-nearest neighbors interactions.  A larger unit cell would enable use to better account for adsorbate interactions; however, it will be shown in Sec. III that lateral interactions between adsorbed protons are minimal so that a $2 \times 1$ unit cell provides a suitable and practical representation of the surface of the RuO$_2$ (110) electrode.  Furthermore, previous experimental and theoretical studies do not indicate any surface restructuring, implying that a minimal $2 \times 1$ unit cell model is acceptable in constructing the Monte Carlo model. A numerical study with varying slab thickness indicates that convergence of the adsorption energies within less than 50 meV is achieved with three slab layers. The supercell is then constructed in such a way that it is symmetric about the center layer as the hydrogens are placed on the surface with a vacuum height of 10 \AA. Identical simulation parameters as the ones employed in the bulk calculations are used for surface simulations with the exception of the Brillouin zone sampling, which is done with an equivalent grid of $2 \times 2 \times 1$ wavevectors.

There are several possible surface terminations with the first being planar and consisting of a mixture of ruthenium and oxygen atoms. The second termination will have bridging oxygen O$_{\rm br}$, located between two ruthenium atoms while leaving the other ruthenium atoms bare. The last possible structure has both on-top oxygen O$_{\rm ot}$ (located on top of the terminal row of ruthenium) and bridging oxygen O$_{\rm br}$, resulting in a fully oxygenated surface. There appears to be debate in the literature regarding the termination of the surface. Some electrochemical measurements point towards a partially oxygenated surface, whereas other results, which more closely represent the electrolytic conditions that are relevant to pseudocapacitor systems, show the surface of the electrode to be completely oxygen-terminated \cite{Knapp2007,Reuter2001, Ozolins2013, Chu2001, Sun2003,Lin2000,Wang2009,Lobo2003,Madhavaram2001}. Our simulations will thus focus on fully oxygenated electrodes, in accordance with previous first-principles studies and references therein \cite{Liu2012, Ozolins2013}. 

\begin{figure}
\centering
\includegraphics[width=\columnwidth]{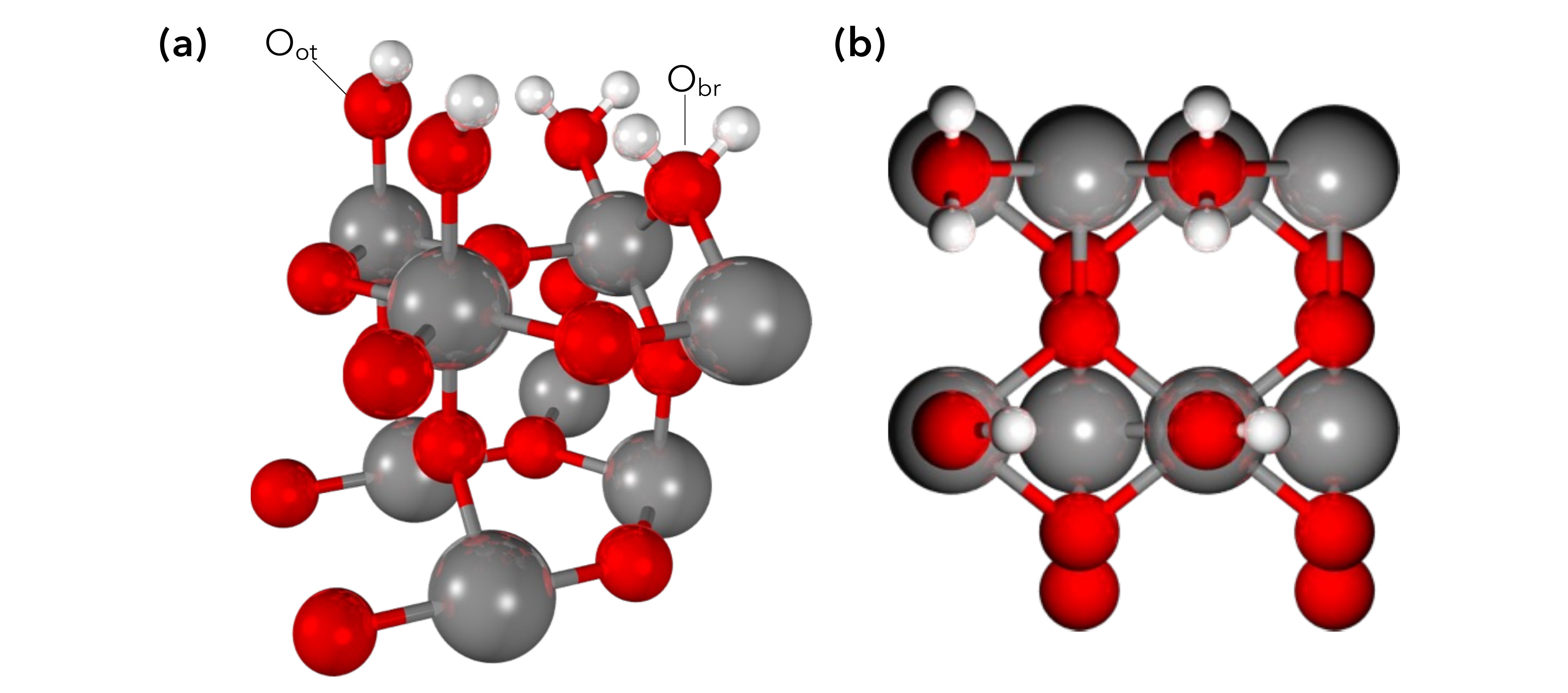}
\caption{(a) Lateral view of RuO$_2$(110) showing the initial locations of surface adsorption sites and (b) top view of the structure.\label{fig:Structure}}
\end{figure}

The hydrogen atoms are then placed on the surface at positions near either O$_{\rm br}$ or O$_{\rm ot}$, in effect creating an OH group that is oriented towards a neighboring oxygen. If placed on an O$_{\rm ot}$, the hydrogen is positioned towards another O$_{\rm ot}$ while the hydrogens near the O$_{\rm br}$ will need to face towards a neighboring O$_{\rm ot}$ as shown in Fig.~\ref{fig:Structure}\ \cite{Ozolins2013}. Using these criteria for placing hydrogen, four different adsorption sites with the possibility of six total hydrogen atoms were considered, which then give 256 admissible configurations in the Monte Carlo calculation. This collection of configurations is then reduced to a more manageable number by making use of surface-symmetry considerations. 

\subsection{Finite-temperature electrochemistry}

The main parameters extracted from the quantum--continuum calculations are the free energy $f_0$ of the hydrogen-covered surfaces, their voltage $\Phi_0$ at zero charge, and their corresponding differential capacitance $C_0$. The voltage is computed by taking the opposite of the Fermi level calculated in DFT, as all of the slab calculations are referenced to the same zero vacuum energy by imposing open boundary conditions at the frontiers of the supercell \cite{Andreussi2014}. In explicit terms, the voltage is related to the Fermi energy through 
\begin{equation}
\Phi_\textup{0} = -\frac{\varepsilon_\textup{F}}{e_\textup{0}},
\end{equation}
where $\varepsilon_{\rm F}$ is the calculated Fermi energy level and $e_0$ is the elementary charge. The free energy for each unit cell can then be computed as
\begin{equation}
f^\alpha=f_{0}^\alpha+\Phi_{0}^\alpha q+\frac{1}{2}\frac{q^2}{C_0^\alpha},
\end{equation}
where $\alpha$ labels the configuration of interest, $f^\alpha$ is the charge-dependent free energy of the supercell, $f^\alpha_0$ is the free energy upon adsorption of the protons at neutral charge, $\Phi_0^\alpha$ is the electrode potential, and $C_0^\alpha$ is the differential double-layer capacitance.  A Helmholtz model is applied to extract the double-layer capacitance.  In this model, the charge of the surface is varied from --10 to 10 $\mu$C/cm$^2$ and an infinitely thin planar countercharge is placed at a distance of 3--5 \AA\ from the surface.  This infinitely thin planar countercharge introduces a discontinuity in the slope of the potential profile, as opposed to a Gaussian planar countercharge for which the change in slope would be gradual.  This model is then exploited to determine the total energy as a function of the surface charge with its second derivative giving the inverse of the double-layer capacitance \cite{Weitzner2017}. The capacitance is found to not significantly vary as the countercharge distance is increased (with changes of 0.5 $\mu$F/cm$^2$ or less upon changing the double-layer thickness from 3 to 5 \AA) due to the diffuse nature of the solvent, which causes the electrostatic potential to be already effectively screened before reaching the Helmholtz plane. The accuracy of the calculated double-layer capacitances is discussed in details in Sec.~\ref{sec:results}.

\begin{table*}[t!]
	\small
	\caption{Surface geometry, configuration number $\alpha$, free energy $f_0^\alpha$, electrode potential $\Phi_0^\alpha$, and differential capacitance $C_0^\alpha$ (at a double-layer thickness of 3 \AA) for the symmetrically inequivalent RuO$_2$(110) surface configurations corresponding to each coverage $\theta$ from 0\% to 150\%. \label{Table-1}}
	\centering
	\begin{tabular*}{0.95\linewidth}{@{\extracolsep{\fill}}lrrrrrr}
	\\
	\hline
	\\
		&\begin{minipage}{0.05\textwidth}
			\centering
			\includegraphics[scale=0.05]{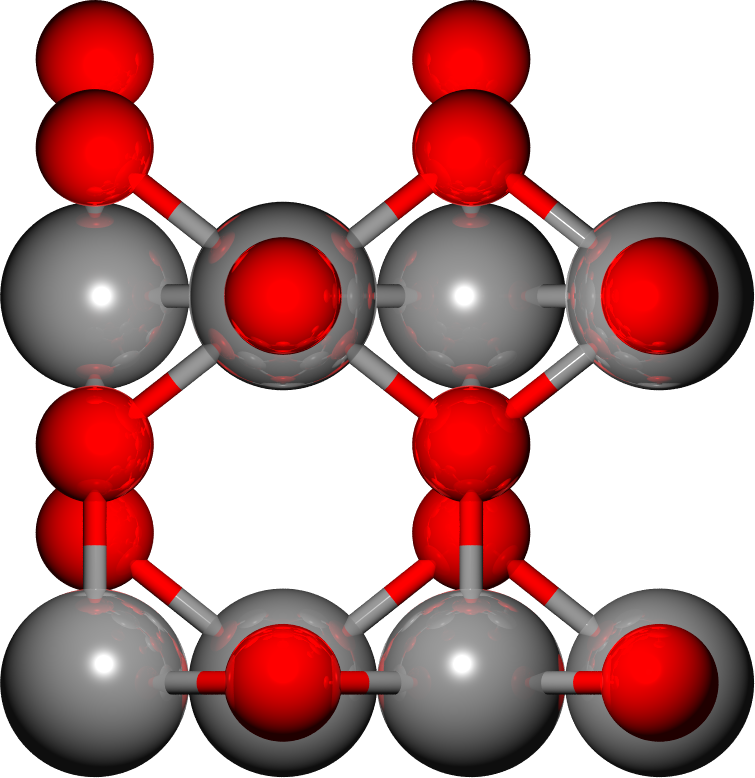}
		\end{minipage} & 
		\begin{minipage}{0.05\textwidth}
			\centering
			\includegraphics[scale=0.05]{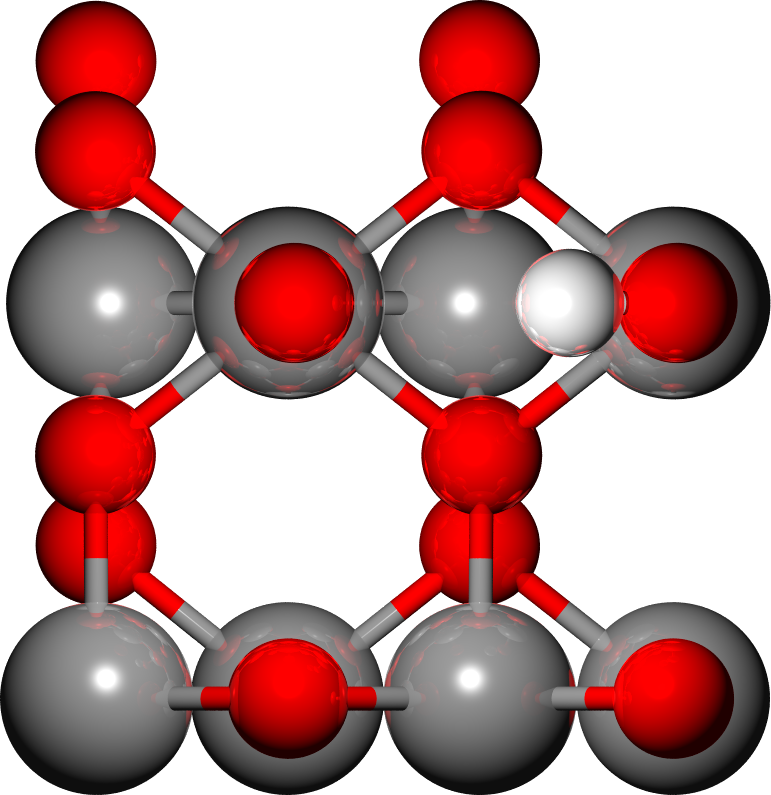}
		\end{minipage}&
		\begin{minipage}{0.05\textwidth}
			\centering
			\includegraphics[scale=0.05]{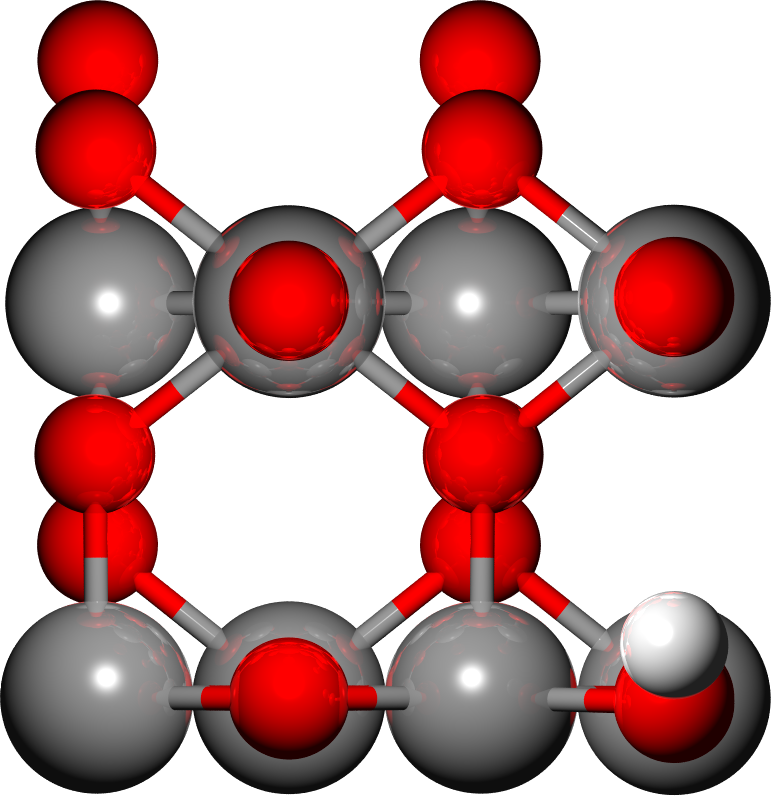}
		\end{minipage} &
		\begin{minipage}{0.05\textwidth}
			\centering
			\includegraphics[scale=0.05]{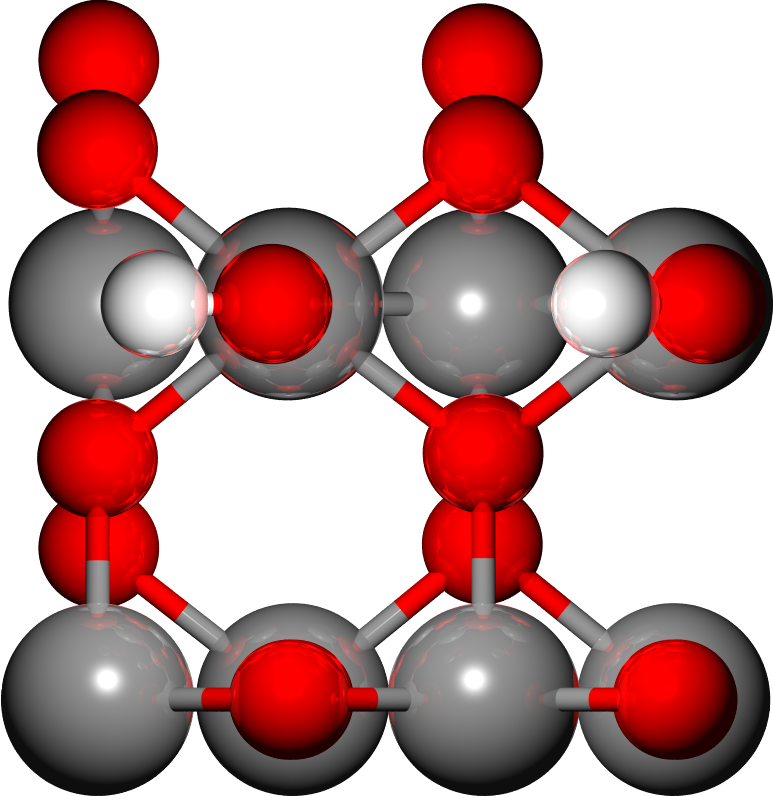}
		\end{minipage} &
		\begin{minipage}{0.05\textwidth}
			\centering
			\includegraphics[scale=0.05]{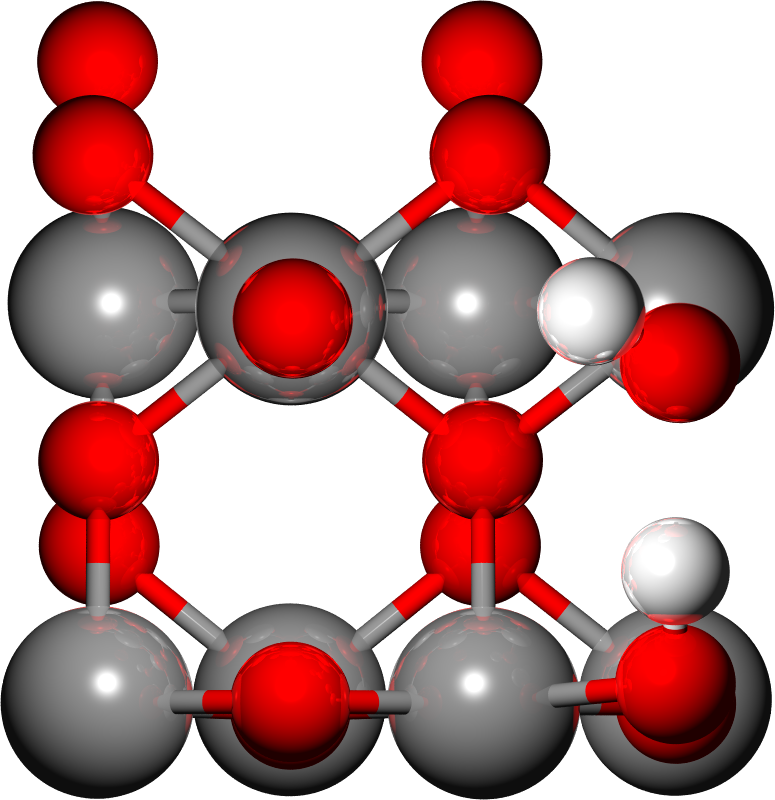}
		\end{minipage} &
		\begin{minipage}{0.05\textwidth}
			\centering
			\includegraphics[scale=0.05]{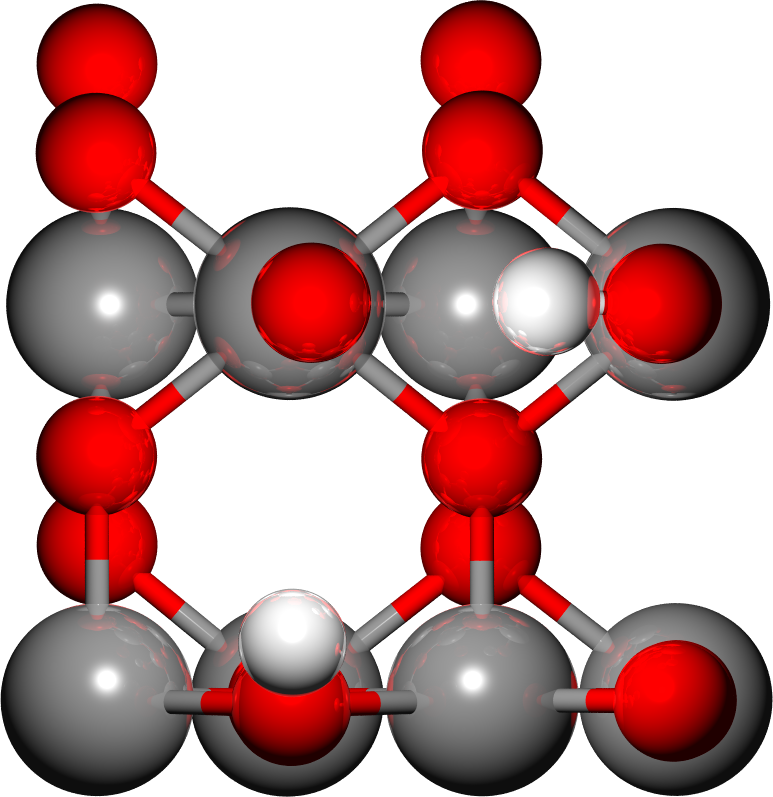}
		\end{minipage}
		\\
		\\
        $\alpha$ & 1 & 2 & 3 & 4 & 5 & 6 \\

		$\theta^\alpha $ & 0\% & 25\% & 25\% & 50\% & 50\% & 50\%  \\
		
		$f_0^\alpha$ (eV)& 0& --2.01& --1.96& --4.07& --3.91& --3.86 \\

		$\Phi_0^\alpha$ (V)& 7.03& 6.69& 6.25& 6.37& 6.34& 6.07 \\

		$C_0^\alpha$ ($\mu$F/cm$^2$)& 7.29& 9.84& 9.49& 10.45& 8.52& 8.22 \\
		\\
		\hline
		\\
		&\begin{minipage}{0.05\textwidth}
			\centering
			\includegraphics[scale=0.05]{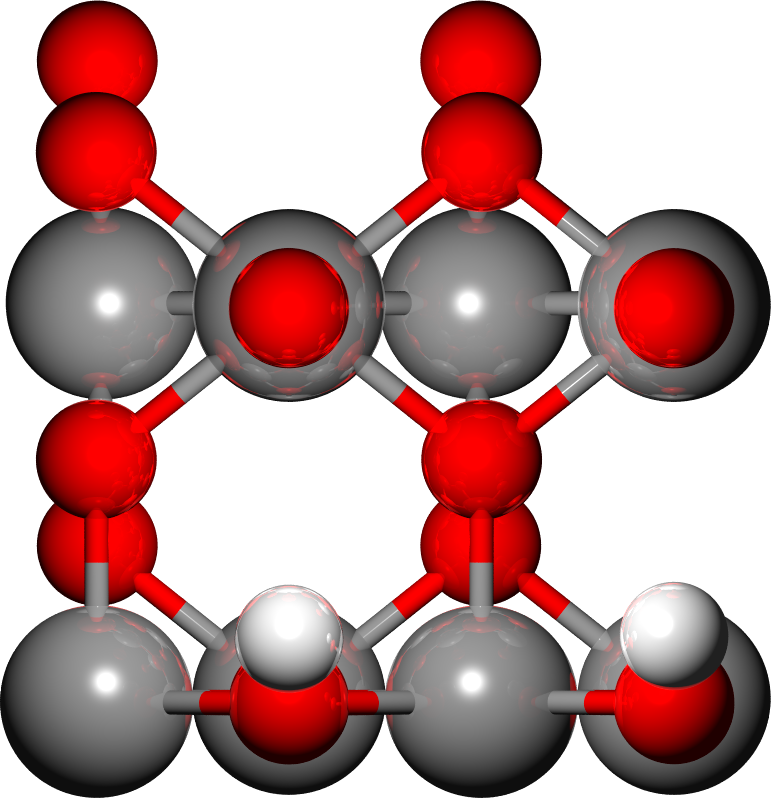}
		\end{minipage} &
		\begin{minipage}{0.05\textwidth}
			\centering
			\includegraphics[scale=0.05]{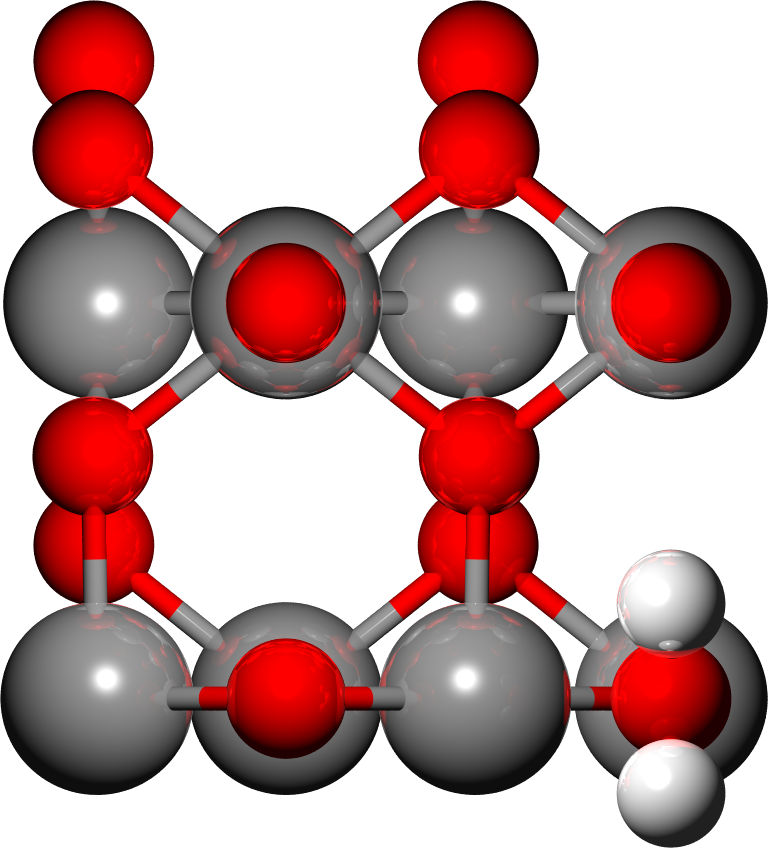}
		\end{minipage} & 
		\begin{minipage}{0.05\textwidth}
			\centering
			\includegraphics[scale=0.05]{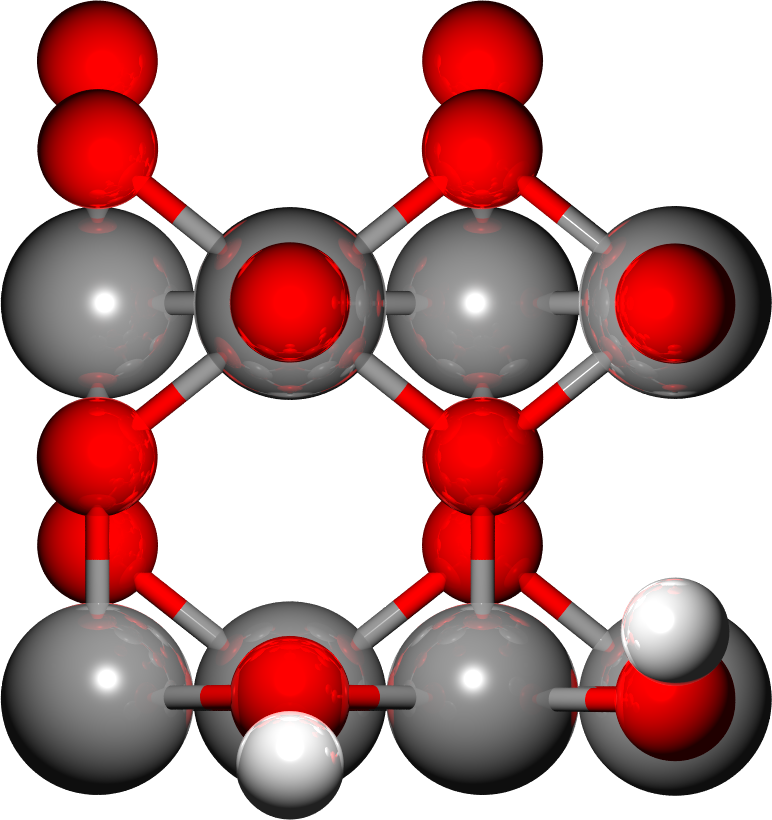}
		\end{minipage}&
		\begin{minipage}{0.05\textwidth}
			\centering
			\includegraphics[scale=0.05]{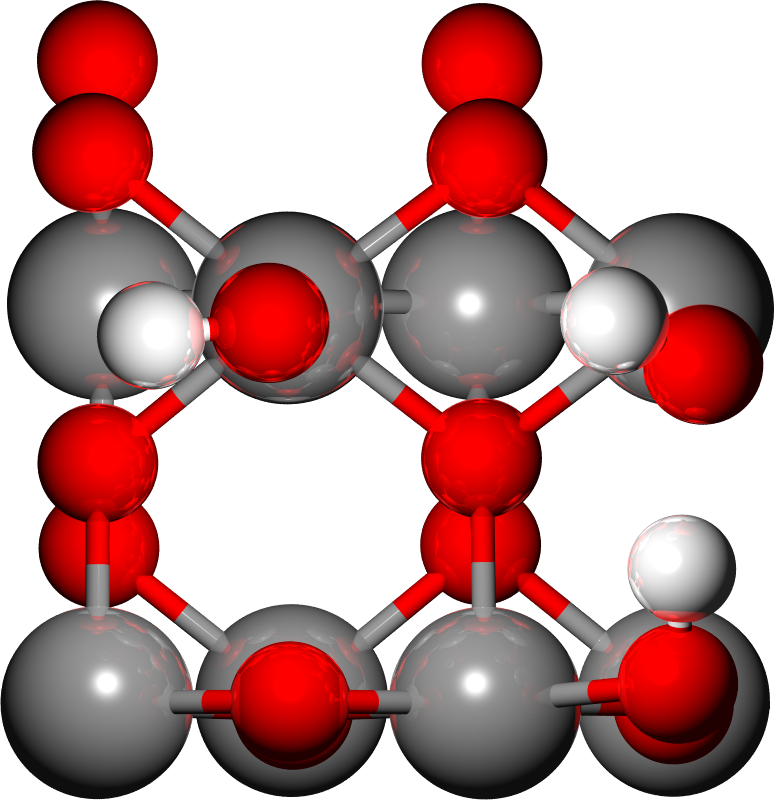}
		\end{minipage} &
		\begin{minipage}{0.05\textwidth}
			\centering
			\includegraphics[scale=0.05]{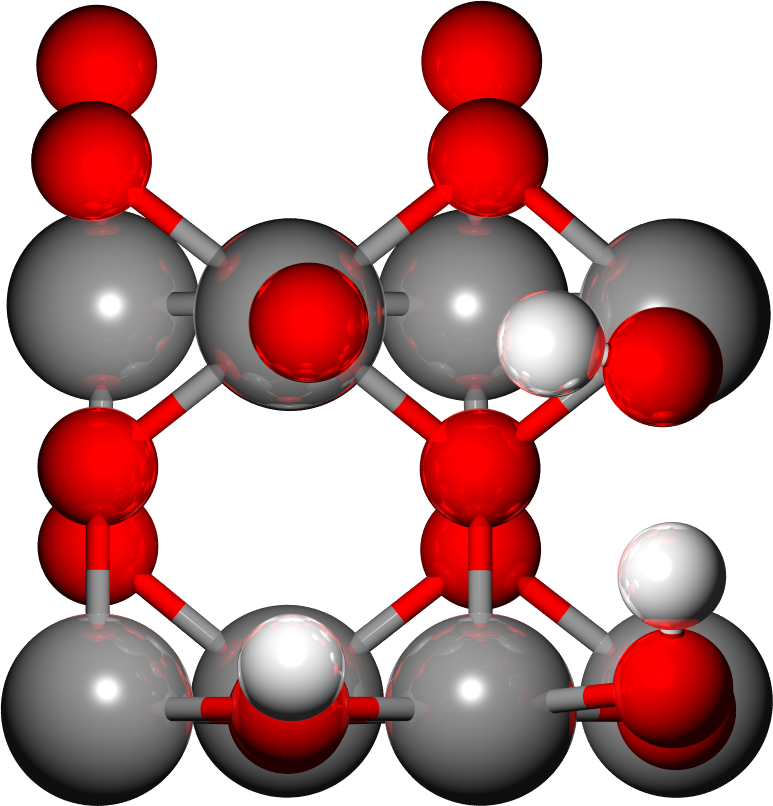}
		\end{minipage} &
		\begin{minipage}{0.05\textwidth}
			\centering
			\includegraphics[scale=0.05]{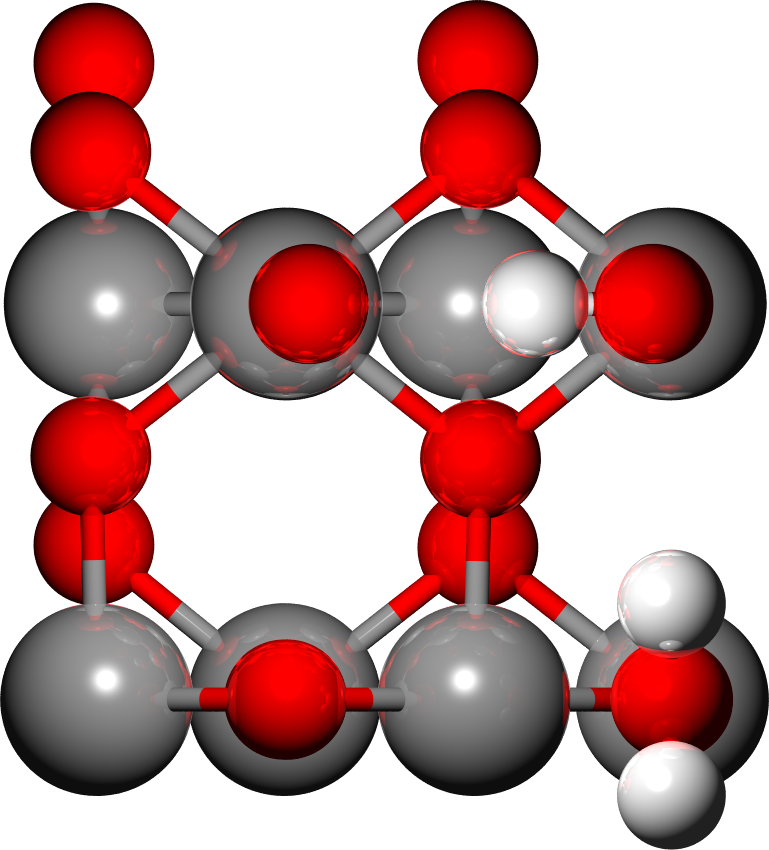}
		\end{minipage} 
		\\
		\\
        $\alpha$ & 7 & 8 & 9 & 10 & 11 & 12 \\

		$\theta^\alpha $ & 50\% & 50\% & 50\% & 75\% & 75\% & 75\%  \\

		$f_0^\alpha$ (eV)& --3.65 & --3.24& --3.64& --5.91& --5.59& --5.17 \\

		$\Phi_0^\alpha$ (V)& 5.67 & 5.33& 5.69& 6.10& 5.77& 5.23 \\

		$C_0^\alpha$ ($\mu$F/cm$^2$)& 11.49 & 8.75& 13.02& 7.98& 9.17& 8.08\\
		\\
		\hline
		\\
		&\begin{minipage}{0.05\textwidth}
			\centering
			\includegraphics[scale=0.05]{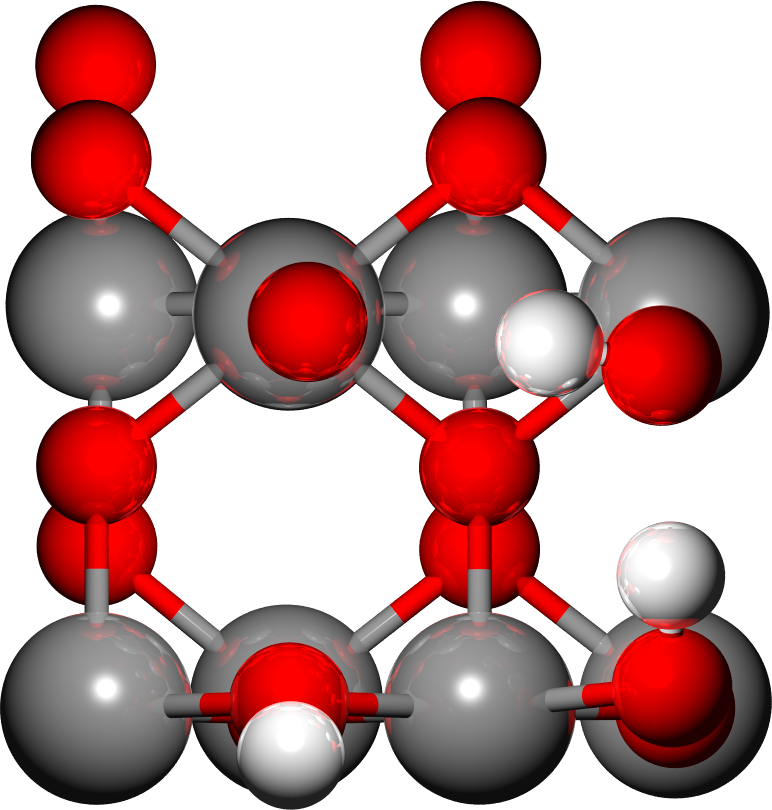}
		\end{minipage} &
		\begin{minipage}{0.05\textwidth}
			\centering
			\includegraphics[scale=0.05]{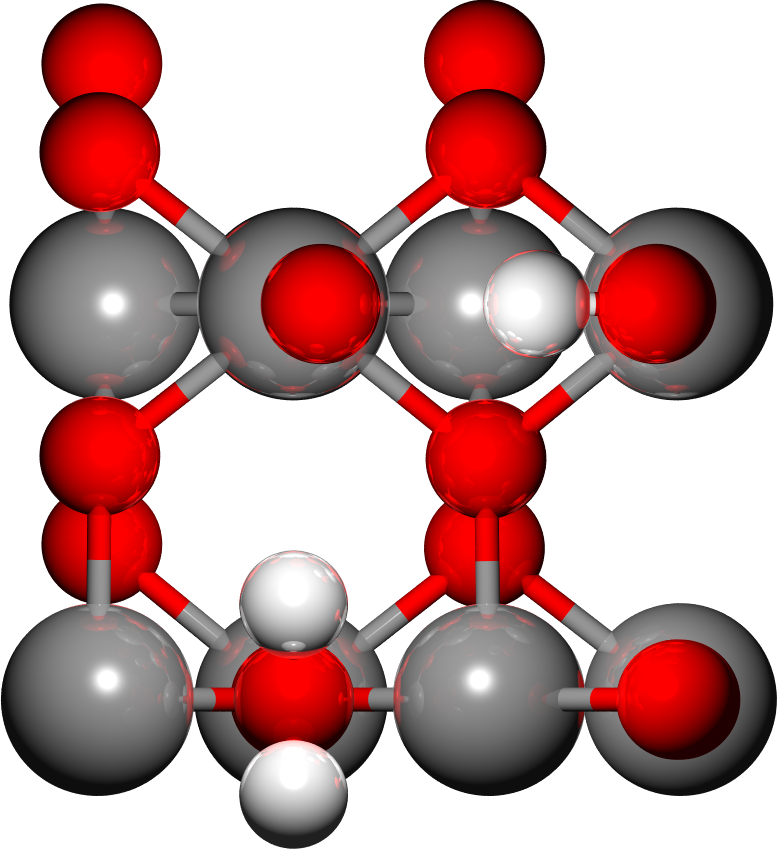}
		\end{minipage} &
		\begin{minipage}{0.05\textwidth}
			\centering
			\includegraphics[scale=0.05]{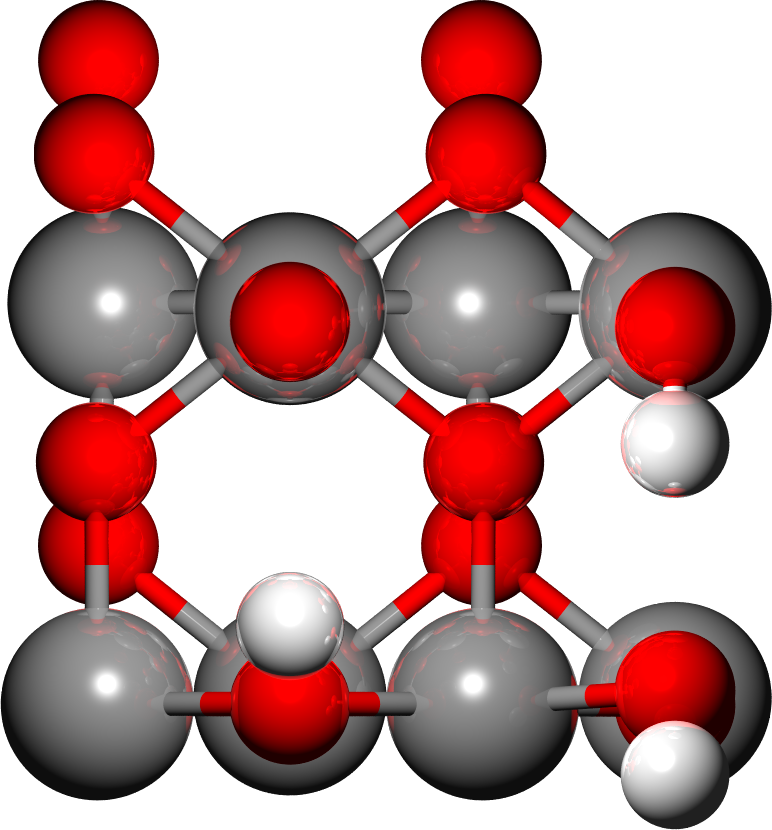}
		\end{minipage} & 
		\begin{minipage}{0.05\textwidth}
			\centering
			\includegraphics[scale=0.05]{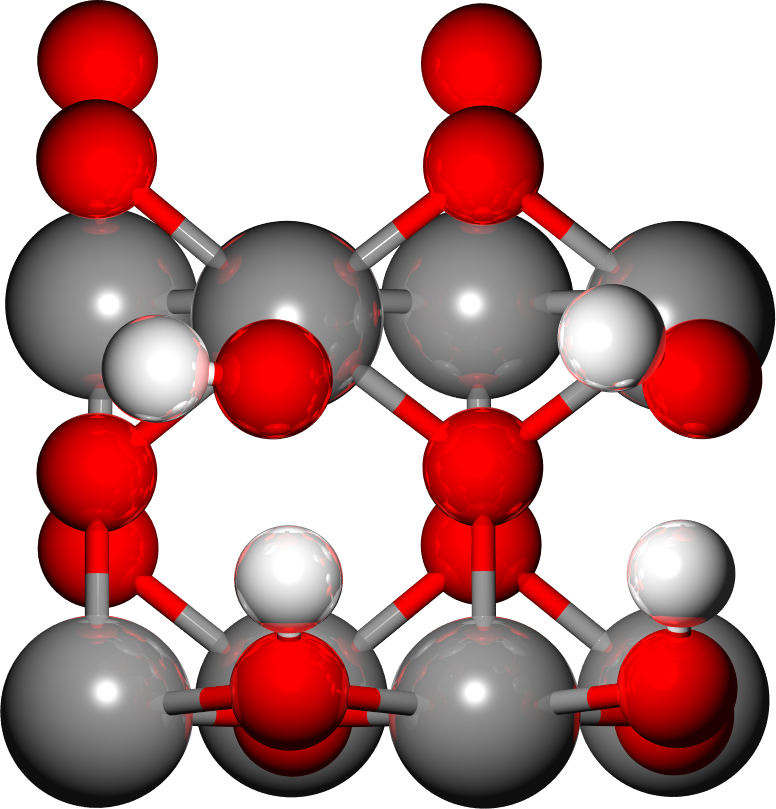}
		\end{minipage}&
		\begin{minipage}{0.05\textwidth}
			\centering
			\includegraphics[scale=0.05]{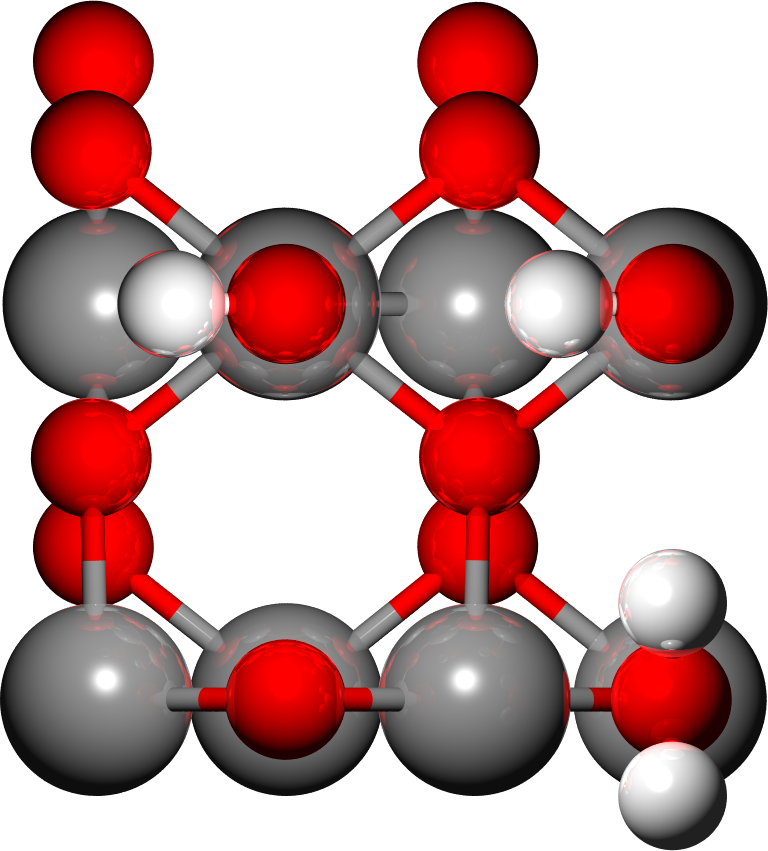}
		\end{minipage} &
		\begin{minipage}{0.05\textwidth}
			\centering
			\includegraphics[scale=0.05]{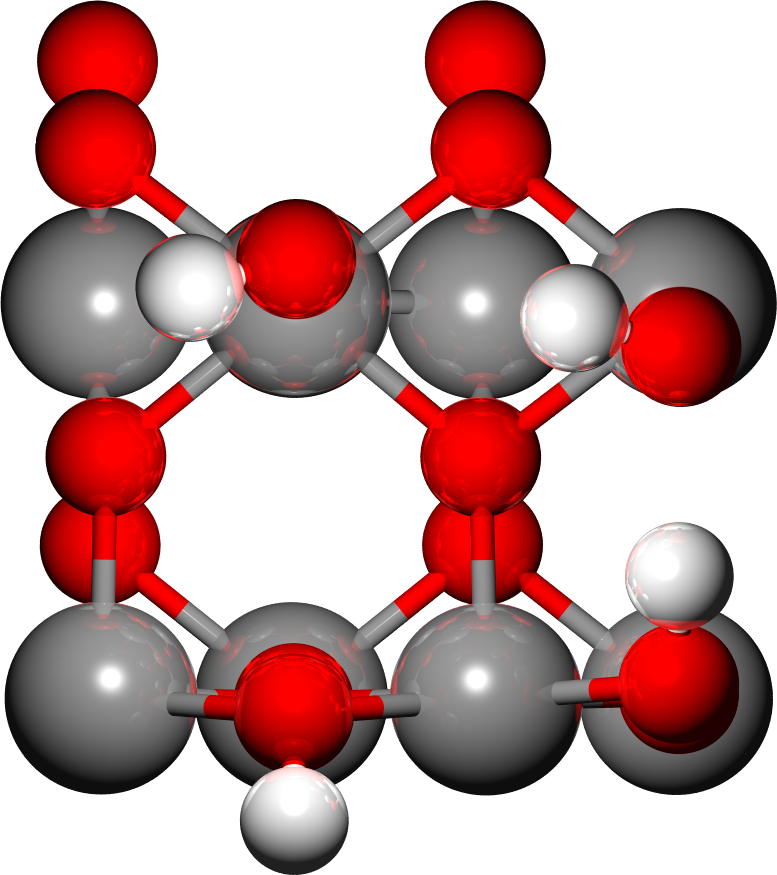}
		\end{minipage} 
		\\
		\\
        $\alpha$ & 13 & 14 & 15 & 16 & 17 & 18\\

		$\theta^\alpha $ & 75\% & 75\% & 75\% & 100\% & 100\% & 100\%  \\

		$f_0^\alpha$ (eV)& --5.60 & --5.04 & --5.04& --7.56& --6.96& --7.47 \\

		$\Phi_0^\alpha$ (V)& 5.78 & 5.17 & 5.53& 5.87& 4.87& 5.78\\

		$C_0^\alpha$ ($\mu$F/cm$^2$)& 9.04 & 8.09 & 9.23& 8.16& 8.01& 9.70\\
		\\
	\hline
	\\
	&\begin{minipage}{0.05\textwidth}
		\centering
		\includegraphics[scale=0.05]{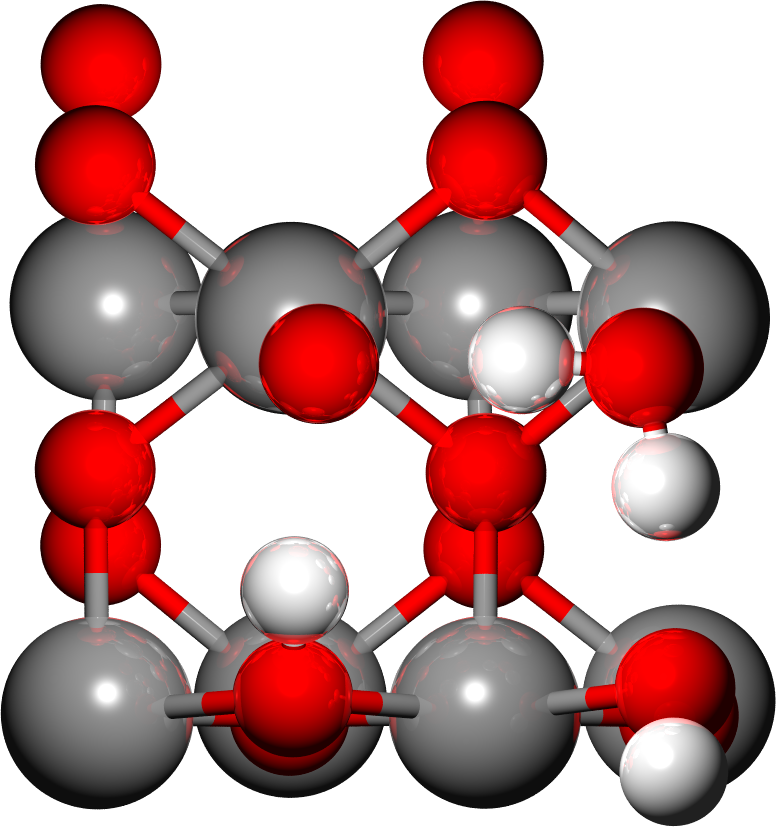}
	\end{minipage} &
	\begin{minipage}{0.05\textwidth}
		\centering
		\includegraphics[scale=0.05]{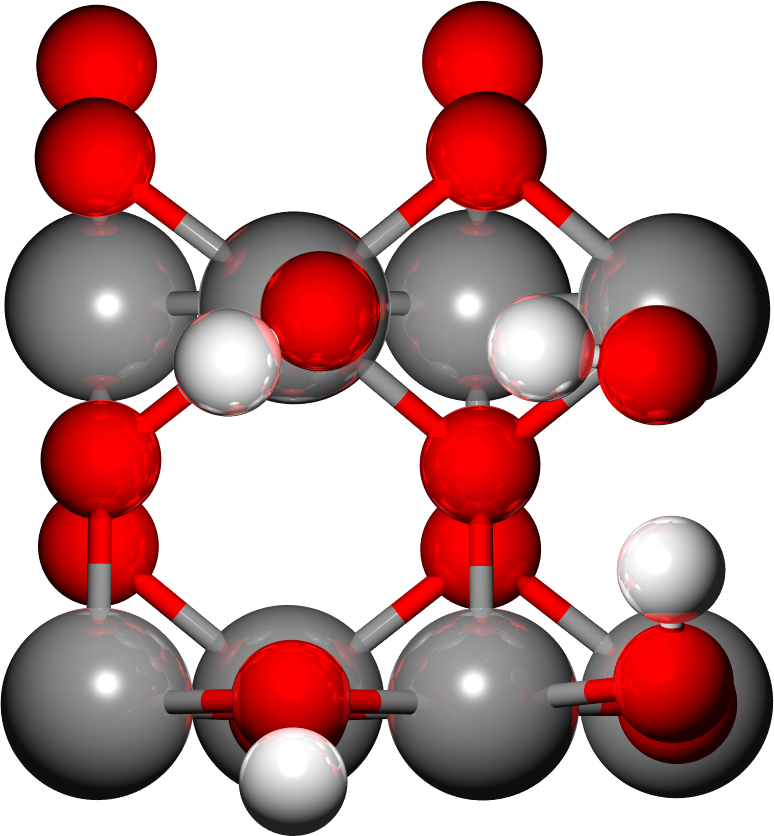}
	\end{minipage} &
	\begin{minipage}{0.05\textwidth}
		\centering
		\includegraphics[scale=0.05]{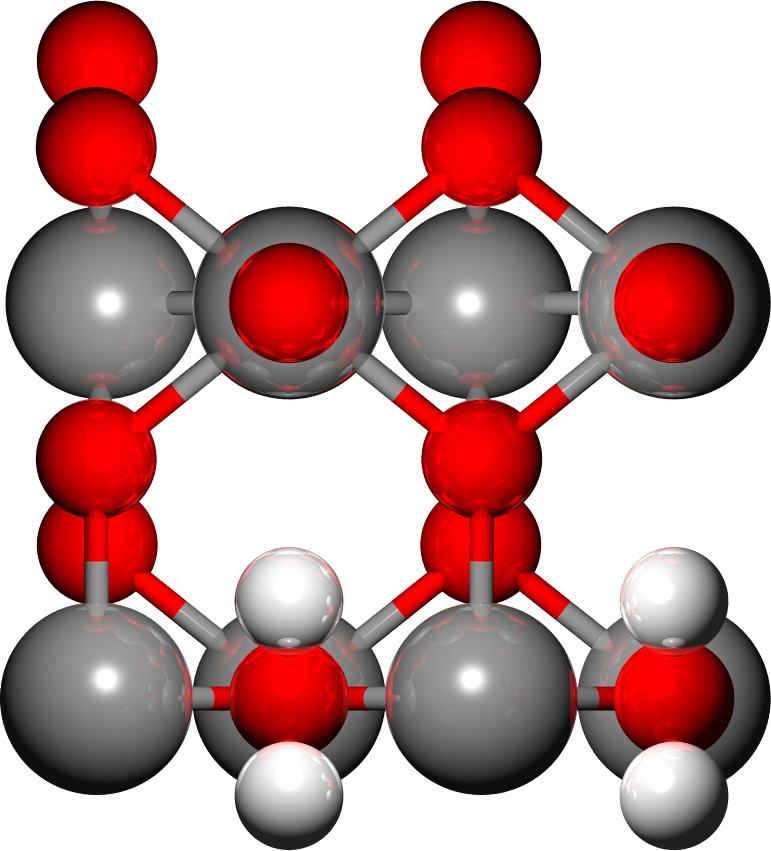}
	\end{minipage}&
	\begin{minipage}{0.05\textwidth}
		\centering
		\includegraphics[scale=0.05]{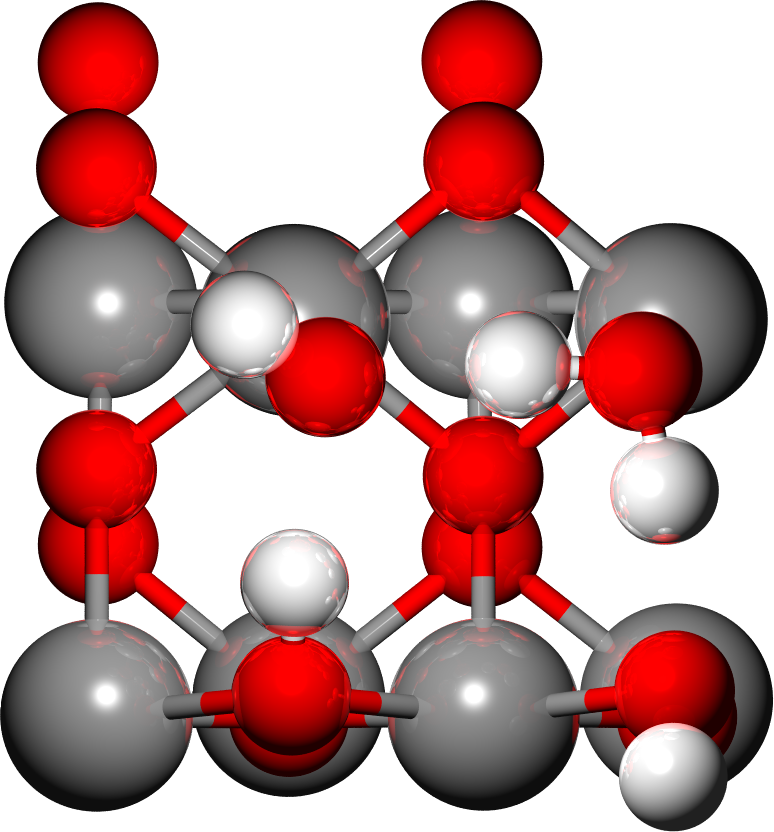}
	\end{minipage} & 
	\begin{minipage}{0.05\textwidth}
		\centering
		\includegraphics[scale=0.05]{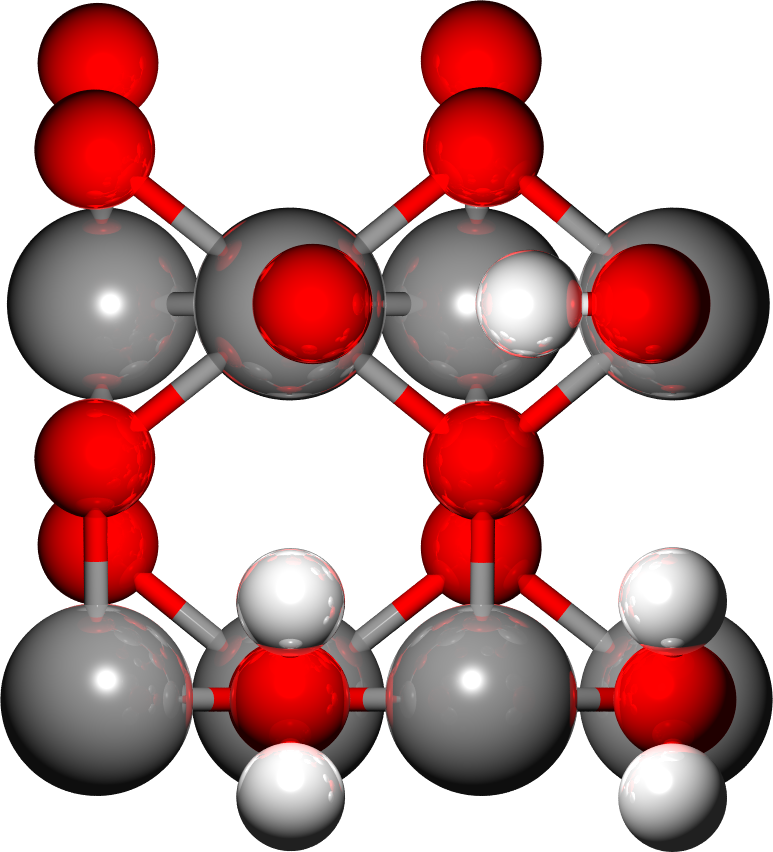}
	\end{minipage}&
	\begin{minipage}{0.05\textwidth}
		\centering
		\includegraphics[scale=0.05]{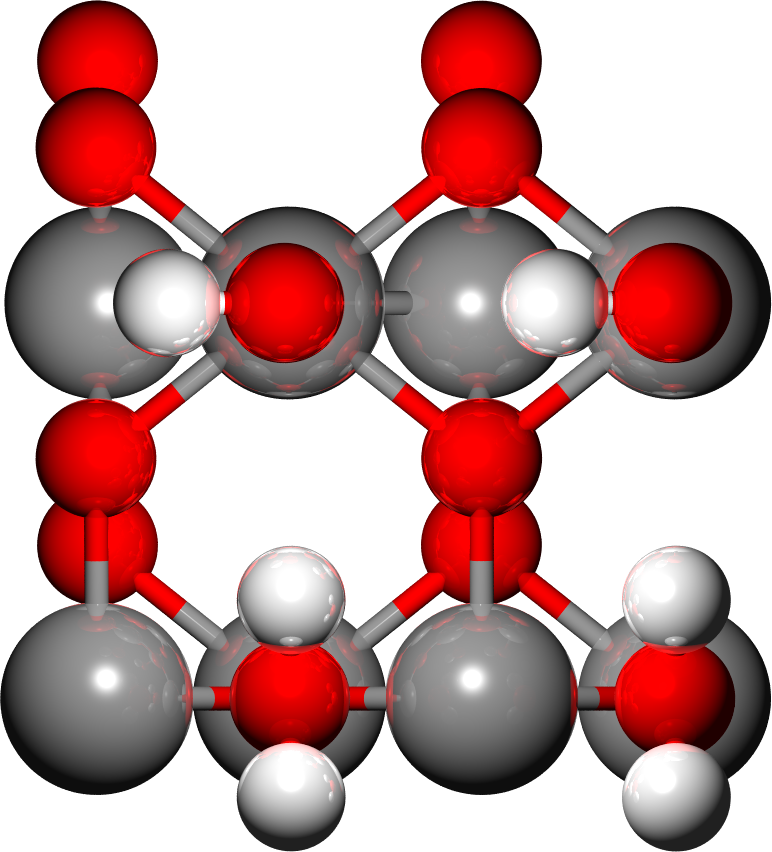}
	\end{minipage}
	\\
	\\
	$\alpha$ & 19 & 20 & 21 & 22 & 23 & 24 \\

	$\theta^\alpha $ & 100\% & 100\% & 100\% &125\% & 125\% & 150\% \\

	$f_0^\alpha$ (eV)& --7.42 & --7.50 & --5.51 & --9.34& --7.34& --9.00 \\

	$\Phi_0^\alpha$ (V)& 5.33 & 5.56 & 3.91 & 4.77& 3.85& 3.48 \\

	$C_0^\alpha$ ($\mu$F/cm$^2$)& 7.63 & 8.43 & 6.53 & 6.60& 6.38& 6.08\\
	\\
	\hline
	\end{tabular*}
\end{table*}

Since these simulations are done in supercell settings, an extrapolation method is used to calculate the chemical potential of the protons on the extended surface of the slab. A robust method to perform this extrapolation consists of averaging the voltage-dependent free energy of the supercells that overlap on a given site.  It can be proved analytically and computationally that this averaging method is at least comparable in accuracy to a second-nearest-neighbor Ising model with pair and triplet interactions, while offering the advantage of being applicable to arbitrarily complex surface configurations.  However, the main requirement in applying this method is to pre-calculate an exhaustive dataset of relevant surface structures, which is computationally feasible for H adsorption on RuO$_2$ and related oxide surfaces.

In explicit terms, the expression of the free energy contribution from a given site reads $\langle f\rangle_j=\frac{1}{4}\sum_{\alpha\in\Lambda_j}f^\alpha$,
where $\langle f\rangle_j$ is the average of the free energy at site $j$ and $\alpha\in\Lambda_j$ represents the sum of all the configurations in the space around the adsorption site, $\Lambda_j$. The same method is applied to calculate the average charge, yielding $\langle q\rangle_j = \frac{1}{4}\sum_{\alpha\in\Lambda_j}q^\alpha$, where $\langle q\rangle_j$ is the average of the charge at site $j$. These equations allow us to express the surface free energy as
\begin{eqnarray}\label{eq:F}
F(\{q^\alpha\})&=&\frac{1}{4}\sum_j \langle f\rangle_j \nonumber \\
 & = &\frac 1{16} \sum_j \sum_{\alpha \in \Lambda_j}  f_{0}^\alpha+\Phi_{0}^\alpha q^\alpha+\frac{1}{2}\frac{(q^\alpha)^2}{C_0^\alpha} 
\end{eqnarray}
and the surface charge as
\begin{equation}
Q(\{q^\alpha\})=\frac{1}{4}\sum_j\langle q\rangle_j = \frac 1{16} \sum_j \sum_{\alpha \in \Lambda_j} q^\alpha.
\end{equation}
Then, by minimizing the free energy $F(\{q^\alpha\})$ at constant overall charge $Q(\{q^\alpha\})$, we obtain the charge in each supercell:
\begin{equation}\label{eq:q}
 q^\alpha (Q)= C_0^\alpha (\Phi (Q)-\Phi_0^\alpha),
\end{equation}
where $\Phi(Q)$ is defined as
\begin{equation}
 \Phi(Q) = \frac{ Q+\frac{1}{16}\sum_j \sum_{\alpha\in\Lambda_j} C_0^\alpha \Phi_0^\alpha}{\frac{1}{16}\sum_j \sum_{\alpha\in\Lambda_j} C_0^\alpha }.
\end{equation}
The charge-dependent electrochemical free energy is thus obtained as
\begin{equation}
\mathscr{F}(Q) = F(\{q^\alpha(Q)\}).
\end{equation}
Using this computational approach, we can calculate the free energy associated to the reaction:
\begin{equation}
* +  {\rm H}^+ + e^- \rightarrow {\rm H}^*,
\end{equation}
where $*$ is the adsorption site on the surface, and $\rm H^*$ is the same site occupied by a proton. The change in chemical potential $\Delta \mu$ for the system can then be represented as
\begin{equation}
\Delta\mu(Q)=\mu_{\rm H^*}(Q)-(\mu_{\textup{H}^+}-e_0\Phi(Q)),
\end{equation}
where $\mu_{\rm H^*}$ is the chemical potential corresponding to the difference between the free energy of the occupied site $\mathscr{F}_{\rm H^*}$ and that of the vacant site, and $\mu_{\rm H^+}$ is the chemical potential of the solvated proton, which is obtained from thermodynamical equilibrium relations as
\begin{equation}
\mu_{\textup{H}^+}=\frac{1}{2}\mu_{\textup{H}_2}^\circ - k_\textup{B}T\ln (10)\textup{pH} + e_0\Phi^\circ_{\textup{H/H}^+},
\end{equation}
where $\mu_{\rm H_2}^\circ$ is the chemical potential of a hydrogen gas molecule and $\Phi^\circ_{\textup{H/H}^+} = 4.4$ V is the standard redox potential for hydrogen. Another equivalent way to solve for the change in chemical potential consists of working at constant potential instead of constant charge, namely,
\begin{equation}
\Delta\mu(\Phi)=\mu_{\rm H^*}(Q(\Phi))-(\mu_{\textup{H}^+}-e_0\Phi),
\end{equation}
where the voltage-dependent charge is defined as
\begin{equation}
Q(\Phi)=\frac{1}{16}\sum_j \sum_{\alpha\in\Lambda_j} C_0^\alpha (\Phi-\Phi_0^\alpha).
\end{equation}

These equations allow for an efficient sampling of the surface configurations under realistic electrochemical conditions based upon the Metropolis acceptance probability $\mathscr{P}=\min(1,\exp({-{\Delta\mu}/{k_{\rm B} T}}))$.  They provide a well controlled, easily implementable, and broadly applicable method to compute the energy cost of the adsorption processes regardless of the complexity of the surface configurations. The results of these simulations are presented in the next section with further discussion on their accuracy.

%%%%%%%%%%%%%%%%%%%%%%%%%%
\section{Results and discussion}
%%%%%%%%%%%%%%%%%%%%%%%%%%

\label{sec:results}

\begin{figure}[b!]
	\centering
	\includegraphics[width=\columnwidth]{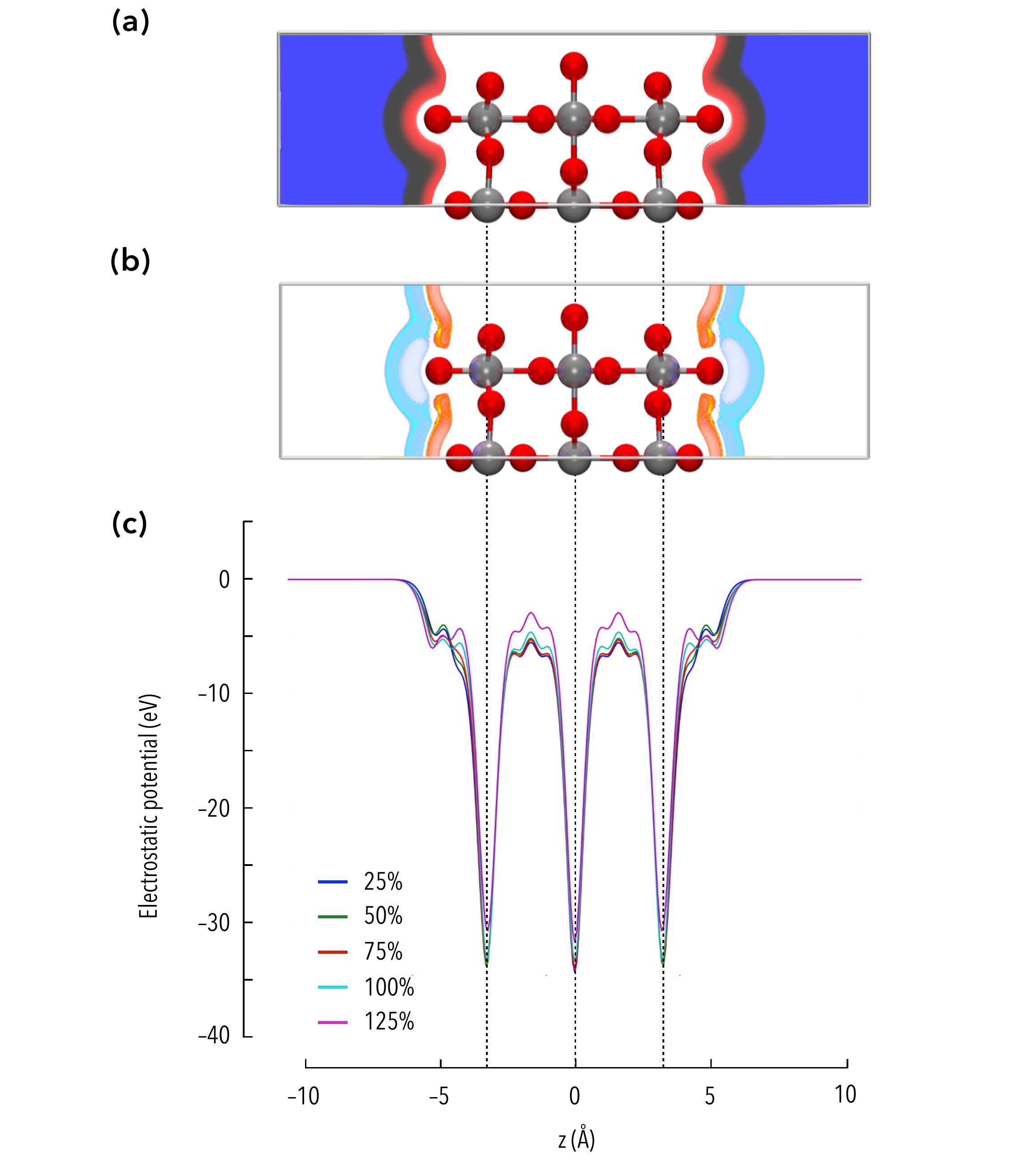}
	\caption{Modeling solvated RuO$_2$(110) electrodes: (a) Self-consistently calculated dielectric constant, (b) the resulting polarization density, and (c) the planar-averaged electrostatic profile for the most stable surface configurations as a function of hydrogen coverage. \label{fig:Potential-Profile}}
\end{figure}

We exploit the above model for the grand-canonical simulation of the voltage-dependent state of the RuO$_2$(110) surface.  We report the calculated equilibrium surface configurations in Table \ref{Table-1} with the corresponding adsorption energies, neutral-electrode potentials, and double-layer capacitances.

The free energy per adsorbed hydrogen is found to vary from --1.38 to --2.03 eV.  Initially, the most stable configurations --- on a per proton basis --- correspond to adsorption at the on-top sites ($\alpha$ = 2--5). Then, as more protons are adsorbed, the bridge sites are progressively occupied and lateral interactions between the protons become significant ($\alpha$ = 21, 23, and 24). Upon adsorption of the sixth hydrogen atom, we observe a sudden increase in the free energy with a shift of the electrode potential below $\Phi^\circ_{\textup{H/H}^+}$, confirming that surface configurations with six hydrogens are unlikely under typical electrochemical conditions \cite{Liu2012,Ozolins2013}. Similar results were obtained by \citeauthor{Liu2012} \cite{Liu2012} where the calculations were done in vacuum. However, the range of values previously obtained is from --1.25 to --1.56 eV per proton, indicating that the inclusion of solvation effects stabilizes proton adsorption significantly.

The corresponding electrostatic-potential profiles are shown in Fig.~\ref{fig:Potential-Profile} for the most stable configurations at each coverage; we observe a gradual shift of the electrostatic potential to higher energies upon increasing the surface coverage. Small perturbations appear near the surface as a result of hydrogen addition and further perturbations occur in the inner layers near the ruthenium atoms. Importantly, we note a marked shift in the potential profile upon adsorbing the fifth hydrogen as a result of a large change in the surface dipole caused by the binding of a second hydrogen at one of the atop sites ($\alpha$ = 22). Accordingly, the Fermi energy undergoes an upward shift, corresponding to a decrease in the potential of zero charge and reflecting the fact that it becomes easier to extract electrons from the surface with every addition of surface hydrogen, which ultimately leads to a destabilization of the high-coverage surface structures upon increasing the voltage.

The calculated double--layer capacitances range from 6 to 13 $\mu$F/cm$^2$. It has recently been suggested by \citeauthor{Montemore2016} \cite{Montemore2016} that the lack of explicit water molecules near the adsorbate layer could overestimate electrostatic screening, which may lead to an underestimation of the capacitance. Notwithstanding this underestimation and despite the recognized simplicity of the Helmholtz model, the predicted capacitances of 6--13 $\mu$F/cm$^2$ are in satisfactory agreement with their experimental counterparts that typically range from 10 to 20 $\mu$F/cm$^2$ \cite{Kotz2000,Lister2002}.

\begin{figure}[t!]
\centering
\includegraphics[width=\columnwidth]{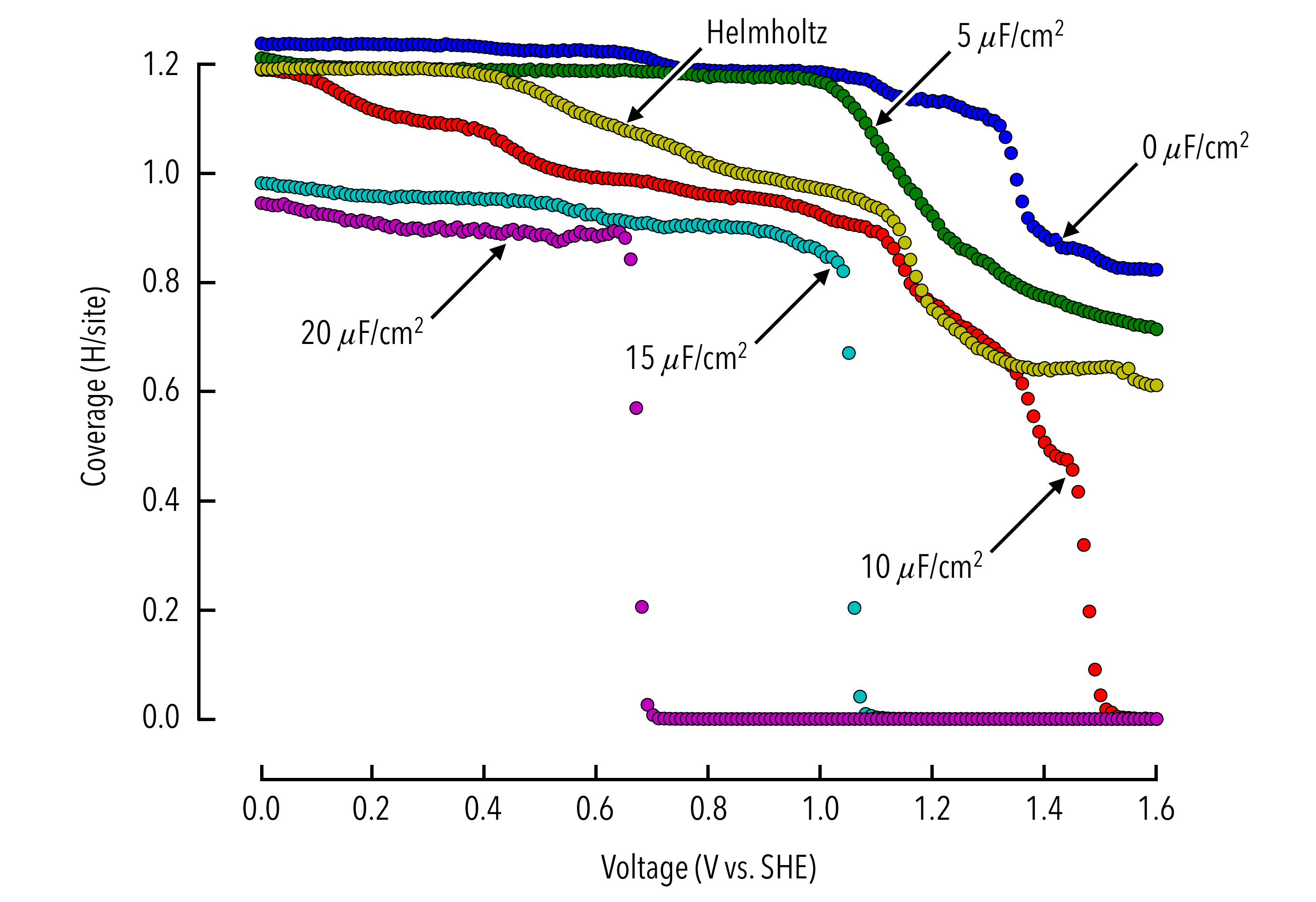}
\caption{Hydrogen-adsorption isotherms obtained by electrochemical Monte Carlo sampling using a fixed double-layer capacitance (from 0 to 20 $\mu$F/cm$^2$) and the full quantum--continuum Helmholtz treatment. \label{Coverage-Voltage}}
\end{figure}

A grand-canonical Monte Carlo model is then parameterized from these results. The Monte Carlo supercell consists of 20 $\times$ 20 adsorption sites. The voltage is varied from 4.4 to 6.0 V in the absolute scale with increments of 0.01 V at a temperature of 300 K. The pH is set at 0.3 to reproduce electrolytic conditions at a concentration of 0.5 M H$_2$SO$_4$ \cite{Lister2002,Ozolins2013}. The sampling is done for more than 100 attempts per site (corresponding to more than 40,000 Monte Carlo moves in total) and each result is averaged over 100 full Monte Carlo runs. The double-layer capacitance calculated directly from the quantum--continuum model is first used for the Monte Carlo sampling.

\begin{figure*}[t]
	\centering
	\includegraphics[width=\textwidth]{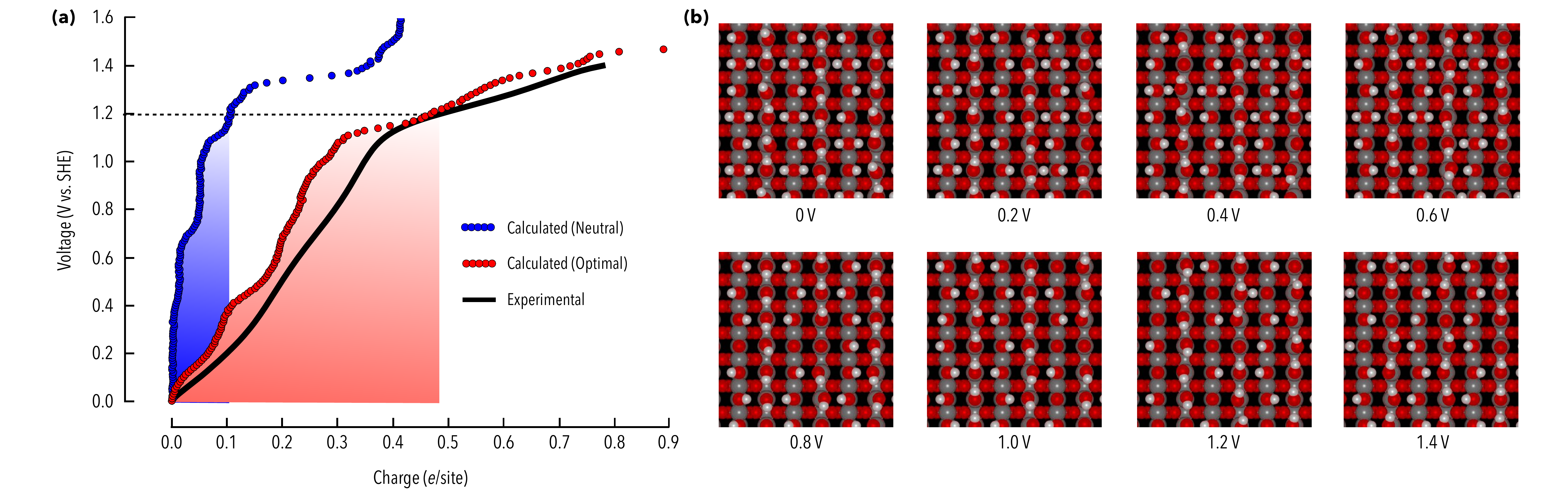}
	\caption{(a) Charge--voltage response of RuO$_2$(110) without surface electrification (corresponding to a double-layer capacitance of 0 $\mu$F/cm$^2$) and including the computationally determined double-layer capacitance that maximizes the overall pseudocapacitance of the electrode (9.5 $\mu$F/cm$^2$), compared with experimental data. The shaded area below the charge--voltage response determines the energy capacity of the electrode (cf.~Fig.~\ref{fig:charge-voltage-response}).  (b) Voltage-dependent evolution of the surface structure under the predicted optimal capacitance. \label{Charge-Voltage}}
\end{figure*}

Applying the Helmholtz model, the hydrogen-adsorption isotherm shows a small plateau region followed by a capacitor-like response, giving a change in surface coverage over an appreciable voltage range. At $\sim$1.1 V there is a sharp transition in the surface coverage before it once again plateaus. As explained previously, the calculated double-layer capacitances are expected to slightly underestimate (by a few $\mu$F/cm$^2$) the initial double-layer capacitance values obtained through the Helmholtz approximation and thus in order to examine the sensitivity of the results with respect to interfacial conditions, the double-layer capacitance is varied up to 20 $\mu$F/cm$^2$, which corresponds to the maximal capacitance that is measured experimentally and matches the upper value obtained from quantum--continuum calculations. Calculated results for different levels of interfacial electrification are shown in Fig.~\ref{Coverage-Voltage}. 

Upon increasing the voltage, we observe the expected decrease in the surface coverage due to the increasingly positive surface charge. However, the nature of this transition is strongly affected by surface electrification. For a system with a double-layer capacitance of 0 $\mu$F/cm$^2$ (that is, in the absence of surface charging), a region of high hydrogen coverage ($\sim$125\%) is observed up to 1.1 V vs.~SHE (standard hydrogen electrode), and high coverages persist even beyond this voltage as a result of the stabilization of the surface configurations provided by the solvent environment. The inclusion of surface electrification is thus critical to capture the voltage-induced desorption of adsorbed hydrogen species in the context of realistic oxide--solvent calculations.  In fact, as the double-layer capacitance is increased, the desorption voltage is shifted to lower redox potentials and the decrease in the surface coverage becomes more significant.  Most of the resulting isotherms exhibit a battery-like response, where constant-coverage plateaus are separated by sharp discontinuities at well-defined values of the voltage.

It is important to note that one of the calculated curves departs markedly from the typical battery-type trends; at a double-layer capacitance of 10 $\mu$F/cm$^2$, we observe a voltage-dependent isotherm showing capacitor-type properties. This capacitor-like response, exhibiting a linear decrease of the surface coverage on a wide voltage window extending up to 1.0 V vs.~SHE, is driven by configurational disorder and is critically dependent on the inclusion of finite temperature and interfacial electrification within the grand-canonical model. Furthermore, it is very encouraging to see that the value of 10 $\mu$F/cm$^2$ is in agreement with the average capacitance of $\sim$9 $\mu$F/cm$^2$ calculated from the quantum--continuum Helmholtz method and that this simple approach captures the occurrence of pseudocapacitive charge storage.

To assess the accuracy of our predictions, the area underneath the experimental RuO$_2$(110) voltammogram is integrated and renormalized by the voltage-sweep rate to obtain the charge--voltage response reported in Fig.~\ref{Charge-Voltage}(a).  It is seen that the first part of the experimental response follows a linear trend as the voltage is increased, which is characteristic of a capacitor, and later exhibits a large change in slope, as expected for a pseudocapacitive system. We then carry out an optimization of the double-layer capacitance to determine the value that maximizes the pseudocapacitive slope; the optimized value of the double-layer capacitance is of 9.5 $\mu$F/cm$^2$. Starting from the same state of charge neutrality and evaluating the area under the predicted curves in Fig.~\ref{Charge-Voltage}(a) up to a voltage of 1.2 V vs.~SHE gives energy densities of 14.3 and 54.6 $\mu$J/cm$^2$ for the 0 $\mu$F/cm$^2$ and 9.5 $\mu$F/cm$^2$, respectively, showing that the double-layer capacitance has a strong influence on the energy-storage capacity of the surface. The calculated optimal capacity is in good agreement with the calculated experimental result of 74.1 $\mu$J/cm$^2$.

Additionally, we note that the model that does not take into account the electrification of the surface (0 $\mu$F/cm$^2$) deviates largely from the experimental response. In contrast, the optimal capacitance model is in close agreement with experiment \cite{Lister2002}; estimating the straight portion of the line under optimal conditions gives a capacitance of 42.6 $\mu$F/cm$^2$ (202.0 F/g), which is consistent with the experimental value of 53.1 $\mu$F/cm$^2$ (251.6 F/g).  Furthermore, the point at which the discontinuity of the slope occurs is calculated to be $\sim$1.1 V vs.~SHE, which is also concordant with the experimental trend and provides a clear indication of the predictive performance of the proposed model. These results offer molecular understanding of the role played by electrolytic conditions in the pseudocapacitive behavior of RuO$_2$, with the ability to capture the influence of the pH, applied voltage, and competition between different proton configurations.

%%%%%%%%%%%%%%%%%%%%%%%%%%
\section{Conclusion}
%%%%%%%%%%%%%%%%%%%%%%%%%%

\label{sec:conclusion}

By performing embedded density-functional theory calculations, we have developed a comprehensive interfacial model to compute the electrochemical properties of oxide--solution interfaces at finite temperature under controlled pH.  We have applied this model for studying pseudocapacitive reactions at RuO$_2$ electrodes, finding qualitative and quantitative agreement with experiment under realistic electrochemical conditions.

The solvent provides a strong stabilization of the adsorbed protons within the electrochemical interface.  This stabilization is compensated by the interaction of the adsorbate-induced dipole with the interfacial electric field. By including these competitive contributions, our simulations highlight the central importance of surface electrification in capturing the electrochemical response of pseudocapacitive electrodes.

Most notably, although the intrinsic double--layer capacitance arising from the interfacial accumulation of electrostatic charges represents a small fraction of the overall electrochemical response of the electrode, it controls to a large extent the onset of the pseudocapacitive reactions and is, therefore, a target of interest in optimizing the electrochemical performance of RuO$_2$ electrodes and other pseudocapacitor materials.  The proposed model provides a robust computational protocol to perform this optimization and assess the maximal performance of new families of pseudocapacitor electrodes.

\section*{Acknowledgments}

The authors acknowledge financial support from Murata Manufacturing and the Center for Dielectric and Piezoelectrics. We thank the Penn State Institute for CyberScience for providing high-performance computing resources and technical assistance throughout this work.

\balance

\bibliography{library}

\end{document}